\documentclass[a4paper,11pt]{article}
\usepackage{pos}

\usepackage{siunitx}

\usepackage{graphicx}
\usepackage{subcaption}
\usepackage{placeins}

\usepackage{hyperref}

\usepackage{braket}

\title{Systematic effects in the lattice calculation of inclusive semileptonic decays}

\author*[a]{R. Kellermann}
\author[b]{A. Barone}
\author[c,d]{A. Elgaziari}
\author[a,e]{S. Hashimoto}
\author[a]{Z. Hu}
\author[c,d,f]{A. J\"uttner}
\author[a,g]{T. Kaneko}
\affiliation[a]{High Energy Accelerator Research Organization (KEK), Ibaraki 305-0801, Japan}
\affiliation[b]{PRISMA+ Cluster of Excellence \& Institut f\"ur Kernphysik, Johannes-Gutenberg-Universit\"at Mainz, D-55099 Mainz, Germany}
\affiliation[c]{School of Physics and Astronomy, University of Southampton, Southampton SO17 1BJ, United Kingdom}
\affiliation[d]{STAG Research Center, University of Southampton, Southampton SO17 1BJ, UK}
\affiliation[e]{School of High Energy Accelerator Science, SOKENDAI (The Graduate University for Advanced Studies), Ibaraki 305-0801, Japan}
\affiliation[f]{CERN, Theoretical Physics Department, Geneva, Switzerland}
\affiliation[g]{Kobayashi-Maskawa Institute for the Origin of Particles and the Universe, Nagoya University, Aichi 464–8602, Japan}

\emailAdd{kelry@post.kek.jp, abarone@uni-mainz.de, A.Elgaziari@soton.ac.uk, shoji.hashimoto@kek.jp, huzhi@post.kek.jp, andreas.juttner@cern.ch, takashi.kaneko@kek.jp}

\abstract{We report on the calculation of the inclusive semileptonic decay of the $D_s$ meson on the lattice. We simulate the $D_s \rightarrow X_s\ell\nu_\ell$ process with M\"obius domain-wall charm and strange quarks, whose masses are approximately tuned to their physical values. Our simulations cover the whole kinematical region. The focus of this work is to present updates on our strategies towards estimating the systematic uncertainties in the determination of the inclusive decay rate. We specifically focus on the systematic errors due to the choice of our approximation strategy and finite-volume effects.}

\FullConference{The 41st International Symposium on Lattice Field Theory (LATTICE2024)\\
 28 July - 3 August 2024\\
Liverpool, UK\\}

\begin{document}
\maketitle

\section{Introduction}

We update on our work to calculate the inclusive semileptonic decay rate of the $D_s$-meson with an emphasis on our progress in estimating systematic uncertainties. Namely, we will focus on the systematic error associated with the Chebyshev approximation of the kernel function \cite{Kellermann:2022mms} as well as the error due to finite-volume effects \cite{Kellermann:2023yec}. For more details on the analysis strategy employed to analyze inclusive semileptonic decay rates in lattice QCD, we refer to \cite{Gambino:2020crt, Gambino:2022dvu, Kellermann:2022mms, Hansen:2017mnd, Hansen:2019idp, Bulava:2021fre}. In particular, the most recent work \cite{Barone:2023tbl} presents a comparison between the Chebyshev and Hansen-Lupo-Tantalo (HLT) approaches to approximate the kernel function in the energy integral.

The remainder of this paper is structured as follows. We present a brief review of the inclusive semileptonic decay on the lattice in Sec.~\ref{sec:InclusiveLattice}. In Sec.~\ref{sec:SystematicError} we discuss the systematic errors of interest in our ongoing work, and our strategies on how to estimate them. The application of our methods is then performed in Sec.~\ref{sec:AnalysisSysErr}. Finally, Sec.~\ref{sec:Conclusion} contains our conclusions.

\section{Inclusive semileptonic decays on the lattice}
\label{sec:InclusiveLattice}

We start by writing the total decay rate of the inclusive semileptonic decay as
\begin{align}
    \Gamma \sim \int_0^{\pmb{q}^2_{\text{max}}} d\pmb{q}^2 \sqrt{\pmb{q}^2} \sum_{l=0}^{2} \bar{X}^{(l)}(\pmb{q}^2) \, ,
\end{align}
where $\bar{X}^{(l)}(\pmb{q}^2)$ contains the integral over the hadronic final-state energy $\omega$
\begin{align}
  \begin{split}
    \bar{X}^{(l)}(\pmb{q}^2) &= \int_{\omega_{\text{min}}}^{\omega_{\text{max}}} d\omega\, W^{\mu\nu}(\pmb{q}, \omega) k_{\mu\nu}^{(l)}(\pmb{q}, \omega) \\
    &= \int_{\omega_{0}}^{\infty} d\omega\, W^{\mu\nu}(\pmb{q}, \omega) K_{\mu\nu}^{(l)}(\pmb{q}, \omega) \, ,
    \label{equ:ContinuumExpression}
  \end{split}
\end{align}
with the hadronic tensor $W^{\mu\nu}(\pmb{q},\omega)$ and a kinematical factor $k_{\mu\nu}^{(l)}(\pmb{q}, \omega)$ depending only on the three-momentum $\pmb{q}$ and the energy $\omega$ of the hadronic final-state. In the second line, we shift the integration limits $\omega_{\text{min}} \rightarrow \omega_0$ and $\omega_{\text{max}} \rightarrow \infty$. While the lower limit $0 \leq \omega_0 \leq \omega_{\text{min}}$ can be freely chosen as there are no states below the lowest-lying energy state $\omega_{\text{min}}$, in order to cut off all contributions above $\omega_{\text{max}}$, we introduce a step function $\theta(\omega_{\text{max}} - \omega)$ which is combined into the \textit{kernel function} $K_{\mu\nu}^{(l)}(\pmb{q}, \omega) = k_{\mu\nu}^{(l)}(\pmb{q}, \omega) \theta(\omega_{\text{max}} - \omega)$. It is defined as
\begin{align}
   K^{(l)}_{\mu\nu, \sigma}(\pmb{q},\omega) = e^{2\omega t_0} \sqrt{\pmb{q}^2}^{2-l} (m_{D_s} - \omega)^l \theta_{\sigma}(m_{D_s} - \sqrt{\pmb{q}^2} - \omega) \, ,
   \label{equ:KernelFunction}
\end{align}
where the sharp cut of the Heaviside function has been replaced by a smooth one by employing a sigmoid function with smearing width $\sigma$. The parameter $t_0$ is introduced to avoid the contact term, which receives contributions from the opposite time ordering corresponding to unphysical states. 

On the lattice we compute four-point correlators $C_{\mu\nu}(\pmb{q}, t)$, which can be related to the hadronic tensor through a Laplace transform
\begin{align}
  \begin{split}
    C_{\mu\nu}(\pmb{q},t) &= \int_0^\infty d\omega \frac{1}{2M_{D_s}} \braket{D_s|\tilde{J}_\mu^\dagger(\pmb{q},0) \delta(\hat{H} -\omega) \tilde{J}_\nu(\pmb{q},0)|D_s} e^{-\omega t} \\
    &= \int_0^\infty d\omega W_{\mu\nu} (\pmb{q}, \omega) e^{-\omega t} \, ,
  \end{split}
  \label{equ:SpectralRepresentation}
\end{align}
where $\tilde{J}^{\nu}(\pmb{q}, 0)$ are the Fourier transformed currents. The above definition corresponds to the spectral representation of $C_{\mu\nu}(\pmb{q}, t)$. By comparing Eqs. \eqref{equ:ContinuumExpression} and \eqref{equ:SpectralRepresentation} it becomes apparent that if the kernel can be approximated using some polynomial of $\exp(-a\omega)$ (where $a$ is set to one for simplicity), lattice correlators can be employed to construct an approximation for $\bar{X}^{(l)}(\pmb{q}^2)$. We thereby reduced the challenge of calculating the inclusive decay rate to one of finding an appropriate polynomial approximation of the kernel function $K^{(l)}_{\mu\nu, \sigma}(\pmb{q},\omega)$.

Our analysis employs the Chebyshev polynomial approach \cite{Gambino:2020crt}. The approximation of the kernel function in terms of shifted Chebyshev polynomials $\tilde{T}_j(x)$ with $x = e^{-\omega}$ is given by
\begin{align}
  \begin{split}
    \braket{K_\sigma^{(l)}}_{\mu\nu} &= \frac{1}{2} \tilde{c}_{\mu\nu,0}^{(l)} \braket{\tilde{T}_0}_{\mu\nu} + \sum_{k=1}^{N} \tilde{c}_{\mu\nu,k}^{(l)} \braket{\tilde{T}_k}_{\mu\nu} \, .
  \end{split}
  \label{equ:KernelApproxCHebyshevMatrix}
\end{align}
Here, $\tilde{c}_{\mu\nu, k}^{(l)}$ are analytically known coefficients and $\braket{\tilde{T}_k}$ are referred to as \textit{Chebyshev matrix elements}. We use the notation $\braket{\cdot} \equiv \braket{\psi^\mu|\cdot|\psi^\nu}/\braket{\psi^\mu|\psi^\nu}$, where $\ket{\psi^{\nu}(\pmb{q})} = e^{-\hat{H} t_0} \tilde{J}^{\nu}(\pmb{q}, 0) \ket{D_s} / \sqrt{2M_{D_s}}$. For simplicity, we skip the indices $\mu, \nu$ in the following.

In practice, the Chebyshev matrix elements are extracted from a fit to the correlator data following
\begin{align}
    \bar{C}(t) = \sum_{j=0}^{t} \tilde{a}_j^{(t)} \braket{\tilde{T}_j} \, ,
    \label{equ:FitCorrelator}
\end{align}
where $\tilde{a}_j^{(t)}$ are obtained from the power representation of the Chebyshev polynomials, see (A.24) and (A.25) of \cite{Barone:2023tbl} for the definition of $\tilde{a}_j^{(t)}$, and $\bar{C}(t)$ is constructed from the correlator as $\bar{C}(t) = C(t+2t_0)/C(2t_0)$. To maximize the available data, we choose $t_0 = 1/2$.
We use priors to ensure that the fitted Chebyshev matrix elements satisfy the condition that the Chebyshev polynomials are bounded, {\it i.e.} $\left|\braket{\tilde{T}_j}\right| \leq 1$. We refer to \cite{Barone:2023tbl} for more details on the Chebyshev approximation and its practical application.

\section{Systematic errors in the inclusive decays}
\label{sec:SystematicError}

We start by introducing the systematic errors of interest in this work. Namely, we analyze the errors introduced by the approximation \cite{Kellermann:2022mms} and due to finite-volume effects \cite{Kellermann:2023yec}. A proper reconstruction of the inclusive decay rate requires the ordered limit
\begin{align}
    \lim_{\sigma\to 0} \lim_{V\to\infty} \bar{X}_{\sigma}(\pmb{q}^2) \,
    \label{equ:OrderedLimits}
\end{align}
of first taking the infinite-volume limit, followed by taking the limit where the smearing of the kernel function is set to zero.

Understanding of finite-volume effects started with the seminal work of L\"uscher \cite{Luscher:1986pf} and is of great importance to many calculations performed on the lattice. The calculation of inclusive decays may suffer from substantial finite-volume effect, as it involves multi-body final states, whose energy levels are discretized by the boundary condition due to the finite volume. Conventionally, the infinite-volume limit is estimated by extrapolating the results from calculations of different volumes. In this work, on the other hand,
we develop a modeling strategy to estimate finite-volume corrections under some assumptions that will be elaborated on in Sec.~\ref{sec:SystematicErrorFV}.

The error due to the approximation is a combination of two effects: first, the smoothing of the kernel function requires the $\sigma \rightarrow 0$ limit, and secondly, the truncation of the Chebyshev approximation at polynomial order $N$ requires the $N\to\infty$ limit. We address this in Sec.~\ref{sec:SystematicErrorApproximation}.

\subsection{Finite-volume effects}
\label{sec:SystematicErrorFV}

The extraction of the spectral density from the correlators $C(t)$ with a finite set of discrete time slices is a well-known ill-posed problem on the lattice. Even assuming that the problem could be solved for a correlator in a finite volume, $C_{V}(t)$, with $V=L^3$ denoting the volume of the lattice, and hence, the spectral density $\rho_V(\omega)$ would be reconstructed, a qualitative difference from its infinite-volume counterpart $\rho(\omega)$ still remains. While the infinite-volume spectral density is a smooth function, $\rho_V(\omega)$ is given by a sum of $\delta$-functions representing allowed states in a finite volume.

The smearing $\sigma$ discussed in Sec.~\ref{sec:InclusiveLattice} allows us to circumvent this problem. By increasing the smearing width $\sigma$, the problem is made arbitrarily mild and the smeared spectral density $\rho_{\sigma, V}(\omega)$ smoothly approaches its infinite-volume counterpart. 
The inclusive decay rate is eventually recovered by taking the limits \eqref{equ:OrderedLimits}.

The finite-volume effects for the spectral density can be sizeable for multi-hadron states, since the allowed states are controlled by the boundary condition. For example, corrections of $\mathcal{O}(1/L^3)$ are expected for the energy spectrum of two-body states. While these would be reduced significantly for the smeared spectral density, their size and scaling in the $V\to\infty$ limit may be non-trivial.
We therefore introduce a model to investigate the volume dependence. 

Among various multi-hadron states, our model concerns the two-body final states, specifically $K\bar{K}$ final states, which are expected to give the dominant contribution. 
Assuming the rest frame, the infinite volume spectral density for the vector current ($J=1$) can be written in the form
\begin{align}
  \rho(\omega) = \frac{1}{64\pi} \omega^2 \left(1 - \frac{4m_K^2}{\omega^2}\right)^{3/2} \, ,
  \label{equ:InfVolSpecVector}
\end{align}
if we ignore interactions between $K$ and $\bar{K}$.
The finite-volume expression is given by
\begin{align}
  \rho_V(\omega) = \frac{\pi}{V} \sum_{\pmb{q}} \frac{\pmb{q}^2}{4(\pmb{q}^2 + m_K^2)} \delta\left(\omega - 2\sqrt{\pmb{q}^2 + m_K^2}\right) \, .
  \label{equ:SpectralDenVec}
\end{align}
In the finite volume, the possible values of $\pmb{q}^2$ for a fixed volume $V$ are given by $\pmb{q} = 2\pi\pmb{l}/L$, where $\pmb{l} = (l_1,l_2,l_3)$ is a vector of integers $l_i$ in $L/2 < l_i \leq L/2$.

In Sec.~\ref{sec:AnalysisSysErr}, we will verify that the model gives a good description of the finite-volume data of inclusive decays, and use it for an extrapolation towards the infinite volume.

\subsection{Finite polynomial approximation}
\label{sec:SystematicErrorApproximation}

\begin{figure}[tb!]
  \centering
  \begin{subfigure}{0.49\textwidth}
    \centering
    \includegraphics[width=\textwidth]{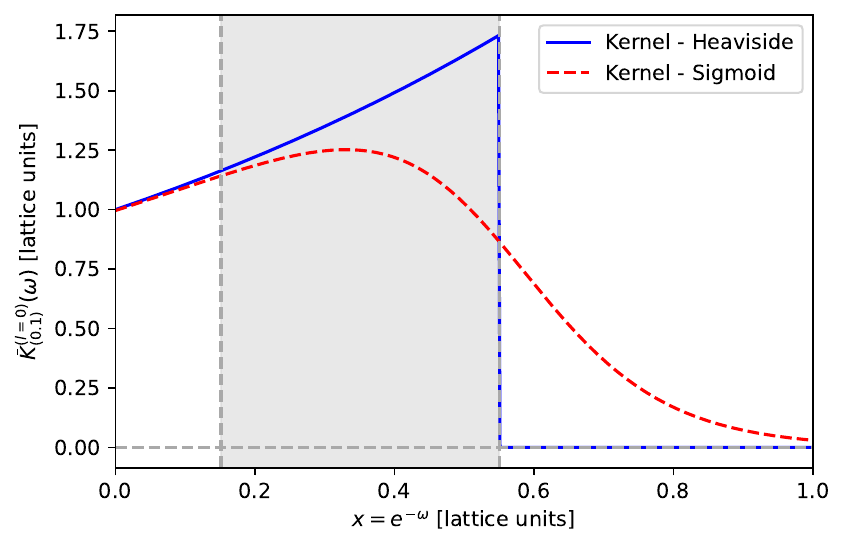}
    \caption{$l = 0$.}
  \end{subfigure}
  \begin{subfigure}{0.49\textwidth}
    \centering
    \includegraphics[width=\textwidth]{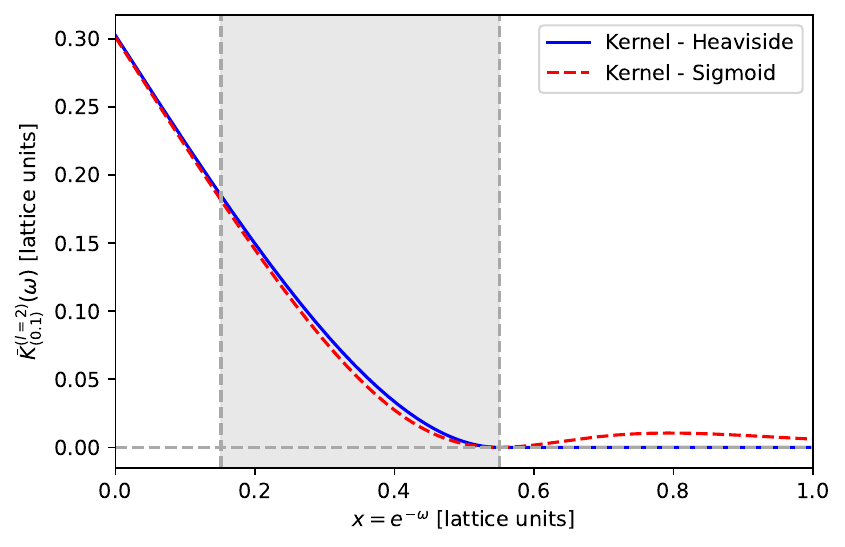}
    \caption{$l=2$.}
  \end{subfigure}
  \caption{Comparison to highlight the differences in the kernel function depending on the choice of the Heaviside or sigmoid function. The Heaviside function in both plots is represented by the solid line, while the sigmoid function uses the dashed line. The smearing width of the sigmoid function is $\sigma=0.1$.}
  \label{fig:CompHeaviSig}
\end{figure}

A first attempt to estimate the systematic error originating from the Chebyshev approximation and the smearing of the kernel function was made in \cite{Kellermann:2022mms}. The two relevant limits are $\sigma \to 0$ and $N \to \infty$. In Fig.~\ref{fig:CompHeaviSig} we visualize the change of the kernel function \eqref{equ:KernelFunction} for $l=0, 2$ before and after smoothing with the sigmoid function.

Our first step is to combine the two required limits by identifying each order of the Chebyshev polynomials as a frequency component of the target function, while interpreting the smearing as the width of the Heaviside function of the kernel. Under these considerations, we introduce the relation between the polynomial order and the smearing as
\begin{align}
  \sigma = \frac{1}{\alpha N} \, ,
  \label{equ:WaveRatioAssumption}
\end{align}
where $\alpha$ is a factor which we set equal to one. In this way, the $N \to \infty$ limit now directly translates to the $\sigma \to 0$ limit.

Our error estimate is then based on the mathematical property that the Chebyshev polynomials are bounded, {\it i.e.} $|\tilde{T}_j(x)| \leq 1$. We redefine our approximation as
\begin{align}
  \bar{K}_{\sigma}^{(l)}(\omega) \simeq \frac{\tilde{c}_0}{2} + \sum_{j=1}^{N_{\text{Cut}}} \tilde{c}_j \tilde{T}_j(e^{-\omega}) + \sum_{\substack{k = \\ N_{\text{Cut}}+1}}^{N} \tilde{c}_k \tilde{T}_j(e^{-\omega}) \, ,
  \label{equ:ApproxErrorEstimate}
\end{align}
where $N_{\text{Cut}}$ is the highest order of the Chebyshev matrix elements that can be properly reconstructed from the data following \eqref{equ:FitCorrelator}. The second term, estimating higher orders, is obtained by repeatedly drawing a random set of Chebyshev matrix elements from a uniform distribution in $[-1,+1]$ and taking the standard deviation of this sample. It only contributes to the error and does not change the central value.

\section{Results}
\label{sec:AnalysisSysErr}

\subsection{Systematic error - Finite-volume corrections}
\label{sec:SysErrFV}

We consider the decomposition of the spectral density
\begin{align}
    \rho(\omega) = \rho_0 \delta(\omega - m_X) + \rho_{\text{Ex}}(\omega) \, ,
    \label{equ:SpectralDeconstruction}
\end{align}
into a ground state contribution and a contribution containing all excited states, specifically focusing on the two-body states defined in \eqref{equ:InfVolSpecVector}. 
We fit the lattice data with
\begin{align}
	C(t) = A_0 e^{-E_0t} + s(L) \sum_{i} A_i e^{-E_i t} F(E_i) \, ,
	\label{equ:ModelFitFunction}
\end{align}
where we pick out the ground state contribution and collect all excited-state contributions into the sum in the second term. The allowed energy $E_i$ is given by the momentum $\pmb{q}$ in \eqref{equ:SpectralDenVec} as $2\sqrt{m_K^2+\pmb{q}^2}$; the corresponding $A_i$ is also inferred from \eqref{equ:SpectralDenVec}. Effectively, $A_i$ determines the relative weights between different energy states normalized to the amplitude of the lowest energy state. 
To obtain a more realistic picture, we include an additional factor $F(E_i)$ motivated by the kaon form factor of the vector-dominance form $F(E) = 1/(E^2 - m_\phi^2)$.
This form factor does not take into account the initial $D_s$, but still provides a reasonable model to describe the process $D_s\to K\bar{K}\ell\nu$ with a free parameter $s(L)$.

\begin{figure}[tb]
	\centering
 	\begin{subfigure}{0.49\textwidth}
   		\centering
    	\includegraphics[width=\textwidth]{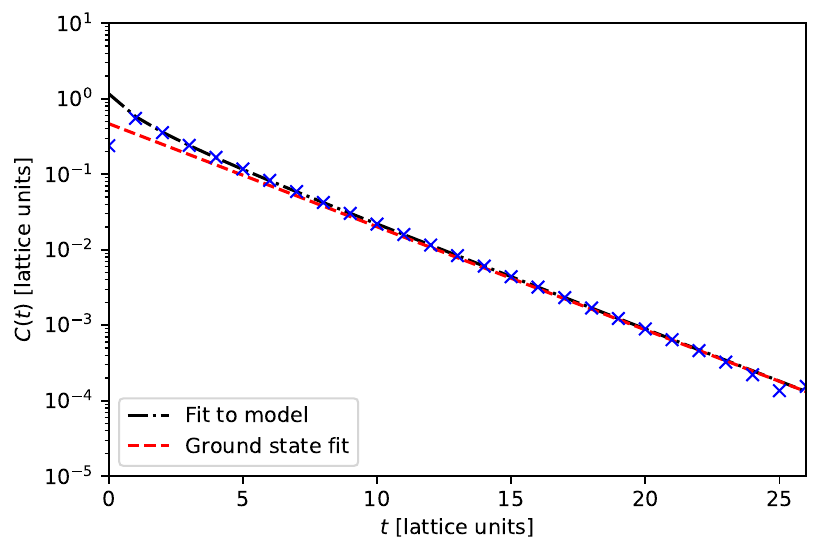}
  	\end{subfigure}
        \begin{subfigure}{0.49\textwidth}
   		\centering
    	\includegraphics[width=\textwidth]{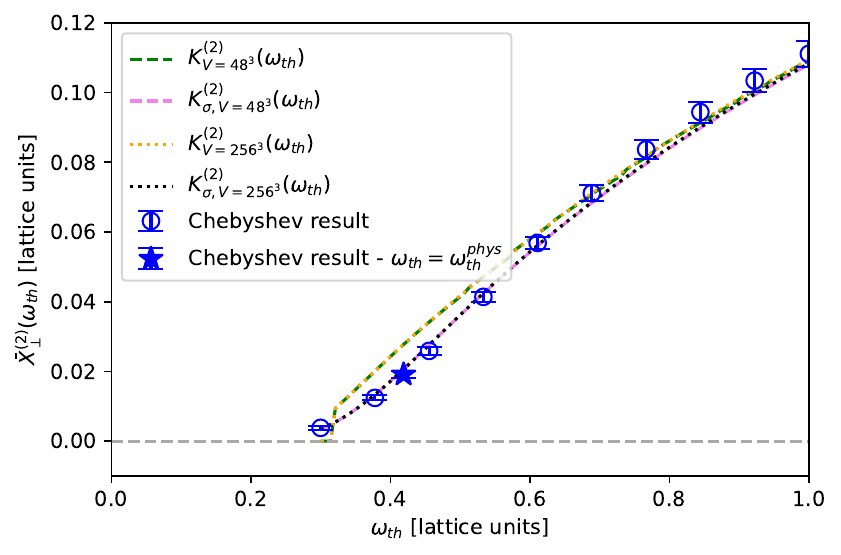}
  	\end{subfigure}
  	\caption{Left: Four-point correlation function for the spatial components $\langle A_i^\dagger(t)A_i(0)\rangle$ with zero momentum insertion.
    The dashed line represents the single-exponential fit to determine the ground state. The black dash-dotted line represents the fit results using the prescription \eqref{equ:ModelFitFunction}. 
    Right: A comparison of the model prediction and the lattice data for the finite-volume corrections. The model prediction is obtained for the smeared and unsmeared kernel, respectively, as a function of the threshold value $\omega_{\text{th}}$ for two volumes $V=48^3$ (dashed) and $256^3$ (dotted). The smeared kernel is denoted with $\sigma$ in the subscript. The lattice data are obtained using the Chebyshev approximation for the full data set. The physical threshold $\omega_{\text{th}}^{\text{phys}}$ is highlighted using a star symbol.}
  	\label{fig:FitToCorrelatorModel}
\end{figure}

As a case study, we consider the contribution of spatial current insertions $A_i^\dagger(t)A_i(0)$ to $X_{AA}^{\parallel}(\pmb{q}^2)$ at zero momentum. In this channel, the lightest hadronic state is the $\phi$ meson, which has $J^P=1^{-1}$. Fig.~\ref{fig:FitToCorrelatorModel} shows the correlator and our estimate of finite-volume corrections. Focusing first on the correlator, we show the single exponential fit to the lattice data as well as the fit result using our model prescription \eqref{equ:ModelFitFunction}. The fit using the model describes the correlator including the short-distance region where excited-state contributions are most prevalent.

We estimate the finite-volume corrections for two volumes $48^3$ and $256^3$. The former corresponds to our lattice data of the physical volume $L\simeq$ 2.6~fm, while the latter is used as a proxy for the infinite-volume limit. The results are shown as a function of the upper limit of the energy integral $\omega_{\text{th}}$. For the physical semileptonic decay process, the upper limit of integration is fixed by $\theta(\omega_{\text{th}}^{\text{phys}}-\omega)$, where $\omega_{\text{th}}^{\text{phys}} = m_{D_s} - \sqrt{\pmb{q}^2}$ is the threshold given by the kinematics. Here, we take advantage of choosing arbitrary values of $\omega_{\text{th}}$ to investigate the validity of our model. We observe a nice agreement between the lattice data (data points) and the model prediction (curves). Within the model predictions, we do not observe any strong dependence on the volume or smearing in the case considered here. This is a consequence of the kernel used in this example, as the spatial currents contribute only with $l=2$ in $X_{AA}^{\parallel}(\pmb{q}^2)$. Comparison with Fig.~\ref{fig:CompHeaviSig} suggests that the kernel smoothly approaches the threshold and no strong dependence on the smearing is observed. This study needs to be repeated for other channels as this behavior is not necessarily expected for all channels.

\subsection{Systematic error - Chebyshev approximation}
\label{sec:SysErrApproximation}

\begin{figure}[!tbp]
    \centering
    \begin{subfigure}{0.49\textwidth}
        \centering
        \includegraphics[width=\textwidth]{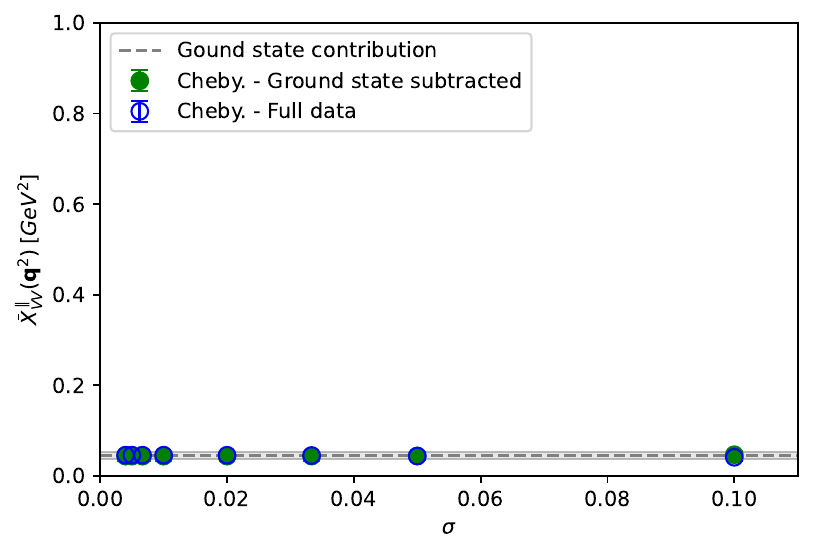}
        \caption{$\pmb{q} = (0,0,0)$}
    \end{subfigure}
    \begin{subfigure}{0.49\textwidth}
        \centering
        \includegraphics[width=\textwidth]{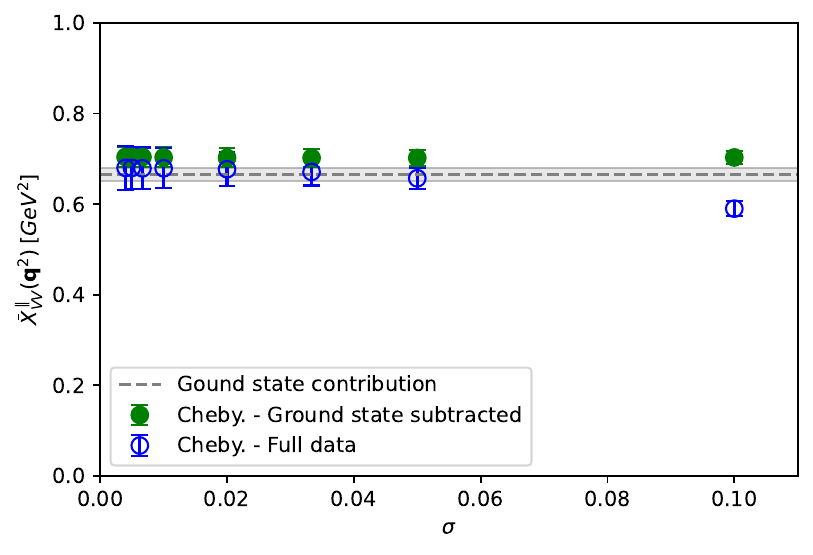}
        \caption{$\pmb{q} = (0,0,1)$}
    \end{subfigure}

    \begin{subfigure}{0.49\textwidth}
        \centering
        \includegraphics[width=\textwidth]{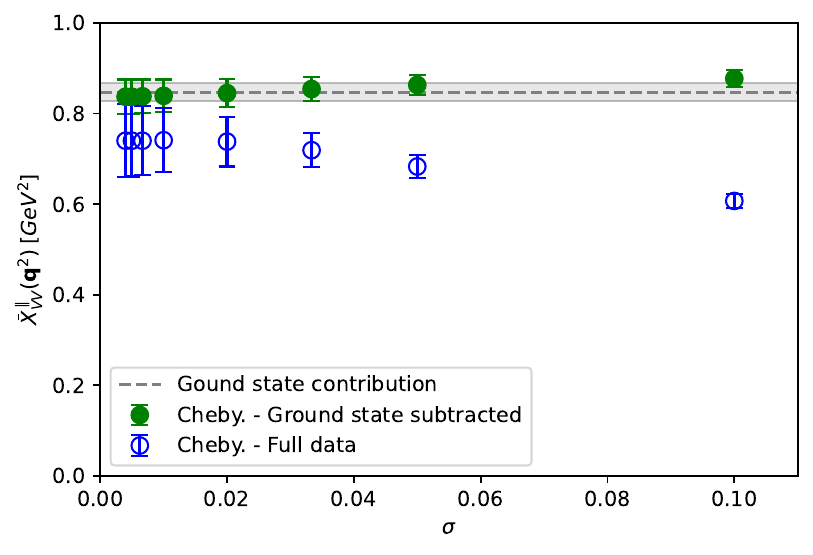}
        \caption{$\pmb{q} = (0,1,1)$}
    \end{subfigure}
    \begin{subfigure}{0.49\textwidth}
        \centering
        \includegraphics[width=\textwidth]{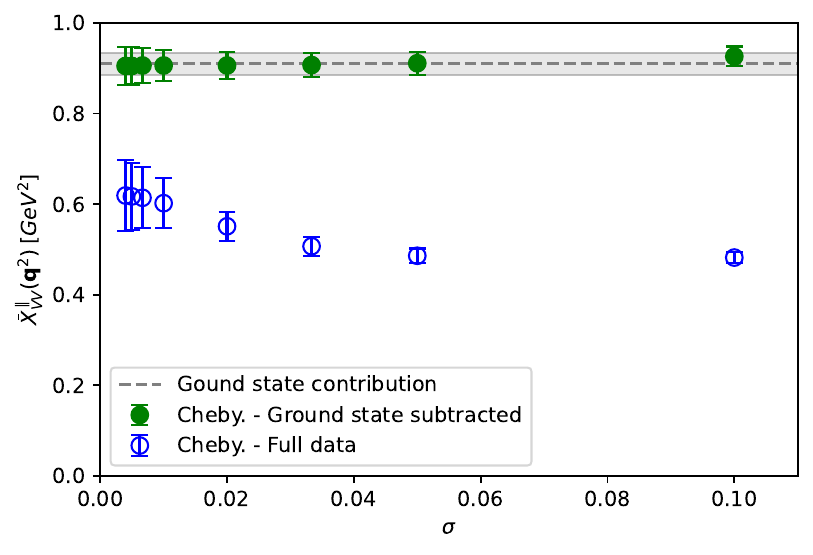}
        \caption{$\pmb{q} = (1,1,1)$}
    \end{subfigure}
    \caption{Errors of $\bar{X}_{VV}^{\parallel}(\pmb{q}^2)$ as a function of $\sigma=1/N$. Estimates are plotted for all values of $\pmb{q}$ used in our simulations.
    The gray band represents the expected ground state contribution. The empty symbols are obtained from applying the Chebyshev approximation on the full data set, while the filled symbols are obtained treating the ground-state contribution exactly and using the Chebyshev approximation only on the remaining correlator. We show the error bars obtained assuming the 1$\sigma$ band using the error estimate for a uniform distribution of the Chebyshev matrix elements.}
  	\label{fig:NExpansion}
\end{figure}

We apply the estimate for the Chebyshev approximation as discussed in Sec.~\ref{sec:SystematicErrorApproximation} to $\bar{X}_{VV}^{\parallel}(\pmb{q}^2)$ for all values of $\pmb{q}$ used in our simulations.
The results are shown in Fig.~\ref{fig:NExpansion} as a function of $\sigma=1/N$. We compare the results for two choices on how the data analysis is applied through the Chebyshev reconstruction. We either use the full data set to construct an approximation or subtract the ground state from the correlator before applying the approximation to the remaining correlator. For the latter choice, the ground-state contribution is treated exactly in the energy integral, while the Chebyshev approximation takes care of the excited-state contributions only. The second approach has the advantage that the ground state can be extracted quite reliably from large time separations, and the overall error is minimized because the ground state is expected to give the leading contribution to the total inclusive rate. To provide a reference value, the estimate under the assumption that only the ground state contributes is also included in the plots.

Without subtracting the ground state (open circles), the estimated truncation error of the Chebyshev polynomials grows rapidly especially for larger momenta, where the phase space becomes narrow. Their central value also drifts significantly as the smearing width gets smaller. Treating the ground state exactly (filled circles), the systematic error remains stable. Indeed, the excited-state contributions are insignificant in this channel, and the results are consistent with those of the ground state (gray band).


\section{Conclusions}
\label{sec:Conclusion}

We investigated the systematic effects in the analysis of the inclusive semileptonic decays. Corrections due to finite volume are modeled to control the infinite-volume extrapolation, and found to be insignificant in our setup. The truncation error of the Chebyshev polynomials was found to be important, especially for large recoil momenta. We can largely reduce the error by treating the ground state exactly and applying the inclusive analysis only to the excited-state contributions.

\section*{Acknowledgments}

The numerical calculations of the JLQCD collaboration were performed on SX-Aurora TSUBASA at the High Energy Accelerator Research Organization (KEK) under its Particle, Nuclear and Astrophysics Simulation Program, as well as on Fugaku through the HPCI System Research Project (Project ID: hp220056).

The works of S.H. and T.K. are supported in part by JSPS KAKENHI Grant Numbers 22H00138 and 21H01085, respectively, and by the Post-K and Fugaku supercomputer project through the Joint Institute for Computational Fundamental Science (JICFuS).

\bibliographystyle{JHEP}
\bibliography{bib_file.bib}

\end{document}